\documentstyle[twocolumn,aps,epsf,floats]{revtex}
\begin{document}

\twocolumn[\hsize\textwidth\columnwidth\hsize\csname %
@twocolumnfalse\endcsname

\title{The resonance peak in cuprate superconductors}
\author{Dirk K. Morr $^{a}$ and David Pines $^{a,b}$}
\address{ $^{a}$ University of Illinois at Urbana-Champaign,  
Loomis Laboratory of Physics, 1110 W. Green St., Urbana, Il, 61801\\
$^{b}$ Center for Materials Science, LANSCE, and Center for Nonlinear Studies, Los Alamos National Laboratory, Los Alamos, NM 87545}
\date{\today}
\maketitle
\begin{abstract}
We pursue the consequences of a theory in which the resonance peak observed in inelastic neutron scattering (INS) experiments on underdoped and optimally doped  YBa$_2$Cu$_3$O$_{6+x}$ compounds arises from a spin-wave excitation. We find that it is heavily damped, and thus almost not observable, in the normal state, but becomes visible in the superconducting state due to the drastic decrease in spin damping.  We show that a spin-fermion model correctly describes the temperature dependence of the peak position for  YBa$_2$Cu$_3$O$_7$, as well as the doping dependence of the peak position
and of the integrated intensity. We explain why no resonance peak has been observed in La$_{2-x}$Sr$_x$CuO$_4$, and make several predictions concerning resonance peaks in other cuprate superconductors.

\end{abstract}
\pacs{PACS:74.20.-q,74.25.Ha,75.40.Gb} 

]

\narrowtext
Recent inelastic neutron scattering (INS) experiments have shown that the sharp magnetic collective mode ("resonance peak"), which was first observed in the superconducting state of YBa$_2$Cu$_3$O$_7$ \cite{Exp1} and only appears in the odd channel, also exists in the underdoped YBa$_2$Cu$_3$O$_{6+x}$ compounds \cite{Dai96,Dai98,Fong97,Kei97,Hay97,Bou97}. 
Several groups find that 
as the doping decreases $\omega_{res}$, the peak frequency decreases \cite{Dai96,Dai98,Fong97,Kei97,Hay97,Bou97}, while both, the peak width in frequency space and its integrated intensity increase \cite{Fong97,Kei97}. In the underdoped systems, a considerably broadened peak at $\omega_{res}$ is also observed in the normal state  \cite{Fong97,Hay97}.
Fong {\it et al.}~also find that $\omega_{res}$ shifts to higher frequencies with decreasing temperature in the superconducting state of YBa$_2$Cu$_3$O$_7$ \cite{Fong96}. 

These new results put tight restrictions on the theoretical scenarios proposed after the discovery of the resonance peak in 
YBa$_2$Cu$_3$O$_7$; these ascribed the resonance peak  to a  final state interaction between the fermionic particles \cite{final}, band structure anomalies \cite{anom}, interlayer tunneling \cite{Yin97}, a new collective mode in the particle-particle channel \cite{SO5}, or a collective spin-wave mode brought about by strong antiferromagnetic correlations \cite{Bar95}. In particular, the observation that  $\omega_{res}$ decreases with decreasing doping, while the superconducting gap $\Delta_{SC}$ is approximately constant, as seen in ARPES \cite{arpes} and tunneling experiments \cite{Ren98} on Bi2212, contradicts scenarios in which $\omega_{res} \approx 2\Delta_{SC}$ \cite{final,anom,Yin97}.

In this communication we use the spin-fermion model to show that the spin-wave scenario, perhaps uniquely,  provides a natural explanation for all  above-cited experimental results. The dispersion  of the spin-wave mode  is 
\begin{equation}
\omega^2_q=\Delta_{sw}^2+c_{sw}^2 ( {\bf q-Q} )^2 \ ,
\label{disp}
\end{equation}
where $\Delta_{sw}$ is the spin-wave gap, $c_{sw}$ is the spin-wave velocity, ${\bf Q}=(\pi,\pi)$; the mode is damped due to its coupling to planar quasi-particles.
In the superconducting state, where the spin damping is minimal, the resonance peak  should always be observable provided 
$\Delta_{sw} < \omega_c \approx 2 \Delta_{SC}$ \cite{com1}. 
For underdoped systems, at temperatures such that ARPES \cite{arpes} experiment show a leading edge gap in the quasiparticle spectrum, the spin damping present at higher temperatures is reduced sufficiently that the resonance mode becomes visible. For optimally doped systems with no leading edge gap, the spin mode is overdamped and invisible.
In contrast to the results of Refs.~\cite{final,anom,Yin97},  the existence of the resonance peak is {\it not} related to the interlayer coupling, and should be observed in single-layer compounds if the superconducting gap is large enough. 
We consider first the two-layer system YBa$_2$Cu$_3$O$_{6+x}$, for which the bonding and anti-bonding tight-binding quasi-particle bands are given by 
\begin{eqnarray}
\epsilon^{\pm}_{\bf k} &=& -2t \Big( \cos(k_x) + \cos(k_y) \Big) \nonumber \\
& & -4t^\prime \cos(k_x) \cos(k_y) \pm t_\perp -\mu \ ,
\end{eqnarray}
where $t, t^\prime$ are the hopping elements between  in-plane nearest and  next-nearest neighbors, respectively, $t_\perp$ is the hopping between nearest-neighbors on different planes, and $\mu$ is the chemical potential.
In a spin-fermion model \cite{sfmodel}, the spin-wave propagator, $\chi$,  is given by 
\begin{equation}
\chi^{-1} = \chi_0^{-1} - {\rm Re}\, \Pi - i \ {\rm Im}\, \Pi \ ,
\label{Dyson}
\end{equation}
where $\chi_0$ is the bare propagator, and $\Pi$ is the irreducible particle-hole bubble.

Since the form of the bare propagator in Eq.(\ref{Dyson}) is model dependent, and thus somewhat arbitrary, we choose a form for $(\chi_0^{-1} - 
{\rm Re}\, \Pi)$ which is the lowest order expansion in momentum and frequency of a hydrodynamical form of Re$\, \chi^{-1}$ \cite{Chu94} and can be shown to reproduce the INS experiments in the normal state of the underdoped YBa$_2$Cu$_3$O$_{6+x}$ compounds \cite{Bou97,Morr98},  
\begin{equation}
\chi_0^{-1} - {\rm Re}\, \Pi = { 1 + \xi^2 ({\bf q} - {\bf Q})^2 - \omega^2/\Delta^2_{sw} \over \alpha \xi^2 } \ ,
\label{chifull}
\end{equation}
where $\xi$ is the magnetic correlation length, $\Delta_{sw}=c_{sw}/\xi$,
and $\alpha$ is an overall constant. We assume that  the form of Eq.(\ref{chifull}) does not change in the superconducting state. Although Dai {\it et al.}~\cite{Dai98} have recently found that for frequencies well below the resonance peak, the peaks in the spin fluctuation spectrum occur at   incommensurate positions, which roughly scale with doping,  we argue  that the use of the commensurate form, Eq.(\ref{chifull}), is justified for the description of the resonance peak since $(a)$  no resonance peak has been observed at incommensurate positions, $(b)$
incommensurate structure and the resonance peak are well separated in frequency, and $(c)$ the use of a commensurate instead of an incommensurate form will only influence low frequency results in the normal state.

With these assumptions, we need only calculate the imaginary part of $\Pi$ which describes the damping brought about by the decay of a spin excitation into a particle-hole pair.  
In the odd channel, Im$\, \Pi$ only includes quasi-particle excitations between the bonding and anti-bonding bands. In the superconducting state we find to lowest order in the spin-fermion coupling $g_{eff}$ (for $\omega>0$)
\begin{eqnarray}
{\rm Im } \, \Pi_{odd} &=& { 3 \pi g^2_{eff} \over 8} \sum_{\bf k} 
\Big(1-n_F(E^+_{\bf k+q})-n_F(E^-_{\bf k}) \Big) \nonumber \\
& & \hspace{-1.0cm} \times 
\Bigg[ 1- { \epsilon^+_{\bf k+q} \epsilon^-_{\bf k} + \Delta_{\bf k+q} 
\Delta_{\bf k} \over E^+_{\bf k+q} E^-_{\bf k} } \Bigg]  
 \delta\Big( \omega-E^-_{\bf k} - E^+_{\bf k+q} \Big)  \nonumber \\
& & \hspace{-1.0cm} + \Big(n_F(E^-_{\bf k})-n_F(E^+_{\bf k+q}) \Big) 
\Bigg[ 1+ { \epsilon^+_{\bf k+q} \epsilon^-_{\bf k} + \Delta_{\bf k+q} 
\Delta_{\bf k} \over E^+_{\bf k+q} E^-_{\bf k} } \Bigg]  \nonumber \\
& & \hspace{-1.0cm} \times \Big\{ 
\delta \Big( \omega+E^-_{\bf k} - E^+_{\bf k+q} \Big) - 
\delta \Big( \omega-E^-_{\bf k} + E^+_{\bf k+q} \Big) \Big\} \ ,
\end{eqnarray}
where $n_F$ is the Fermi function, and
\begin{equation}
E^\pm_{\bf k} = \sqrt{ (\epsilon^\pm_{\bf k})^2 + |\Delta_{\bf k}|^2} 
\end{equation}
is the dispersion of the bonding and antibonding bands in the superconducting state, which we assume is described by the d-wave gap
\begin{equation} 
\Delta_{\bf k}=\Delta_{SC} \   { \cos(k_x) - \cos(k_y) \over 2} \ .
\end{equation}
The normal state result for Im$\, \Pi_{odd}$ is recovered with $\Delta_{SC}=0$. 
We determine  $g_{eff}$ by requiring that Im$\, \Pi_{odd}$ reproduces the spin-damping seen at low frequencies in NMR experiments in the normal state of  YBa$_2$Cu$_3$O$_{6+x}$ \cite{Zha96} and assume  that $g_{eff}$ does not change in the superconducting state.

\begin{figure} [t]
\begin{center}
\leavevmode
\epsfxsize=7.5cm
\epsffile{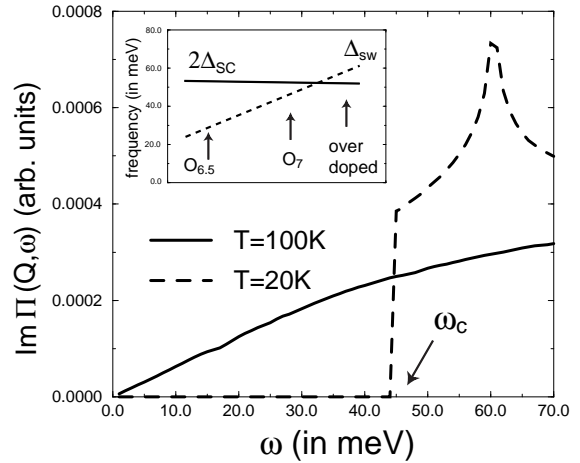}
\end{center}
\caption{ Im$\, \Pi_{odd}$ at ${\bf Q}=(\pi,\pi)$ as a function of frequency in the normal state (solid line) and the superconducting state (dashed line). Inset: Schematic doping dependence of $\Delta_{sw}$ and $2 \Delta_{SC}$.}
\label{damp}
\end{figure}

We first discuss our results for YBa$_2$Cu$_3$O$_7$, which were obtained with the following parameters: $t=300$ meV, $t^\prime=-0.40t$, $t_\perp=0.3t$,  $\mu=-1.27t$ (which corresponds to a $22\%$ hole concentration in the planes), and $\Delta_{SC}(T=0)\approx 25$ meV, a value extracted from the tunneling experiments of Maggio-Aprile {\it et al.}\cite{Mag96}.
 In Fig.~\ref{damp} we present  our calculated result for Im$\, \Pi_{odd}$ at ${\bf Q}=(\pi,\pi)$ as a function of frequency for the normal (solid line) and superconducting  state
 (dashed line). 
In the normal state the spin-damping increase linearly with frequency, as is to be expected \cite{Chu97}. The spin-damping in the superconducting state is characterized by a step-like feature at $\omega_c \approx 2 \Delta_{SC}$ which arises from the creation of a particle-particle pair above the superconducting gap. 
At $T=0$, there are no quasi-particle excitations below $\omega_c$, and  the spin-damping vanishes. The step in Im$\, \Pi$ will then, via the Kramers-Kronig relation, lead to a logarithmic divergence in Re$\, \Pi$. 
This is neglected in Eq.(\ref{chifull}), since the inclusion of fermion lifetime effects as found, e.g., in strong coupling scenarios, eliminates both, the sharp step in Im$\, \Pi$ and the divergence in Re$\, \Pi$. 

\begin{figure} [t]
\begin{center}
\leavevmode
\epsfxsize=7.5cm
\epsffile{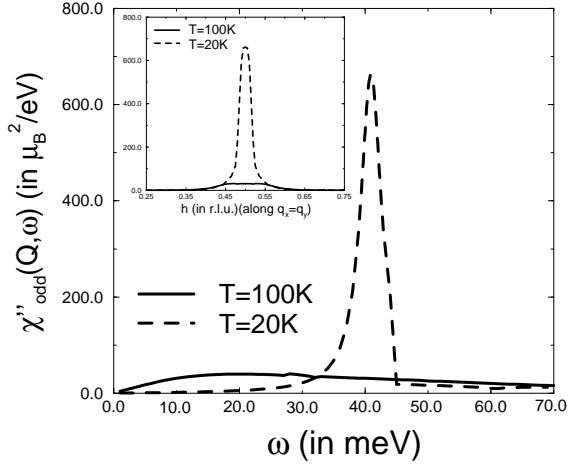}
\end{center}
\caption{ $\chi''$ at ${\bf Q}=(\pi,\pi)$ as a function of frequency in the normal state (solid line) and the superconducting state (dashed line). Inset: $\chi''$ for fixed frequency $\omega=41$ meV along $q_x=q_y$.}
\label{chi}
\end{figure}

In Fig.~\ref{chi} we present our results for  $\chi_{odd}''$ at ${ \bf Q}=(\pi,\pi)$ as a function of frequency for the normal (solid line) and superconducting state (dashed line). 
 In the normal state, the spin excitations are clearly overdamped and $\chi''$ exhibits a flat maximum around $\omega=20$ meV.  
 In contrast, in the superconducting state, the spin-damping is strongly reduced so that the spin-wave mode, which is now very sharp in frequency,  becomes visible at $\omega_{res} = \Delta_{sw}$, which we have chosen to be  $41$ meV to reproduce the experimentally measured peak.
The resulting spin-wave velocity, $c_{sw}=\xi \Delta_{sw} \approx 90$ meV is quite reasonable, since $c_{sw}=(120 \pm 20) $ meV in
YBa$_2$Cu$_3$O$_{6.5}$ and $c_{sw}=180$ meV in the undoped compound \cite{Bou97}.
Our results can be easily understood using Eq.(\ref{chifull}), since the intensity at $\omega=\Delta_{sw}$ is given by
\begin{equation}
\chi''_{odd}({\bf Q}, \omega=\Delta_{sw})= \Big\{ {\rm Im}\, \Pi_{odd} ({\bf Q}, \Delta_{sw}) \Big\}^{-1} \ .
\label{resint}
\end{equation}
If $\Delta_{sw}<\omega_c$, Im$\, \Pi_{odd}$ in the superconducting state is much smaller than in the normal state and consequently, $\chi''_{odd}$ is strongly enhanced at $\omega=\Delta_{sw}$. Our result is robust against changes in the band parameters, in contrast to those scenarios \cite{final,anom,Yin97}, which require a fine-tuning of parameters to observe a resonance peak.

In the inset in Fig.~\ref{chi} we plot $\chi''_{odd}$ for fixed frequency
$\omega=41$ meV along the $(0,0)$ to $(2\pi,2\pi)$ direction in the normal (solid line) and superconducting state (dashed line). We find that the resonance peak is sharp in momentum space, in agreement with the experimental observations \cite{Dai98,Fong96}. 
The sharpness of the resonance peak in both momentum and frequency 
space can be easily understood from Eqs.(\ref{disp}) and (\ref{chifull}). One expects to find a resonance peak at $\omega_q$ which follows the dispersion of the spin excitations, Eq.(\ref{disp}), as long as $\omega_q<2\Delta_{SC}$. 
Since in YBa$_2$Cu$_3$O$_7$, $\Delta_{sw} \approx 2 \Delta_{SC}$ the resonance peak is necessarily confined in both momentum and frequency space. We will show below that this situation is different in the underdoped compounds. Since $\Delta_{sw} \approx 2\Delta_{SC}$ for YBa$_2$Cu$_3$O$_7$, we expect $\Delta_{sw} > 2\Delta_{SC}$ for the magnetically overdoped compounds, where $\xi<2$, and no resonance peak should be observed (see inset in Fig.~\ref{damp}).

Fong {\it et al.}~\cite{Fong96} find that $\omega_{res}$ increases slightly between $T_c$ and $T=10$ K. Since $\Delta_{sw}=c_{sw}/\xi$  
in our model, 
their result implies that $\xi$ decreases over this temperature range, a finding consistent with the changes in the transverse relaxation rate, $T_{2G}^{-1} \sim \xi$, in the superconducting state of  YBa$_2$Cu$_3$O$_7$ measured by Milling {\it et al.}~\cite{Mil98} who report a decrease of $T_{2G}^{-1}$ by about 10 \% with decreasing temperature. Put another way, in our model, the results of Milling {\it et al.}~ require that $\omega_{res}$ increase by some 10\% below $T_c$.

The fact that $\omega_{res}$ is determined by the spin-gap $\Delta_{sw}$ explains the absence of the resonance peak in the even channel of
 YBa$_2$Cu$_3$O$_{6+x}$. For YBa$_2$Cu$_3$O$_{6.5}$, 
Bourges {\it et al.}~\cite{Bou97} find a spin-gap, $\Delta_{sw}^{even} = 53$, meV in the even channel  which may be expected to be  larger in YBa$_2$Cu$_3$O$_7$. Since $\Delta_{sw}^{even} >\omega_c$ the damping of the even spin excitations around ${\bf Q}=(\pi,\pi)$ will {\it not} decrease upon entering the superconducting state and no resonance peak is to be expected.

\begin{figure} [t]
\begin{center}
\leavevmode
\epsfxsize=7.5cm
\epsffile{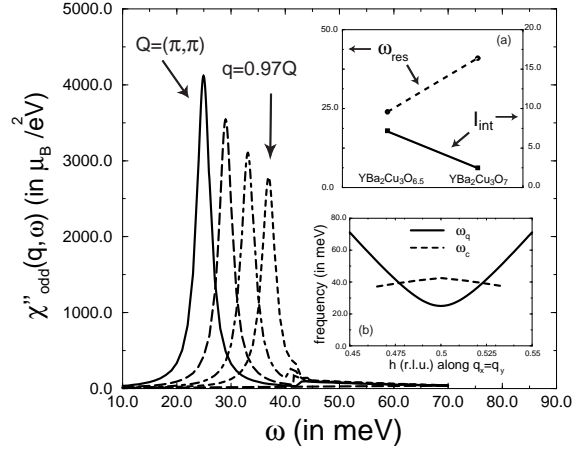}
\end{center}
\caption{ The resonance peak for momenta in the vicinity of ${\bf Q}=(\pi,\pi)$. Inset $(a)$: The doping dependence of $\omega_{res}$ (in meV, dashed line) and I$_{int}$ (in $\mu_B^2$, solid line). Inset $(b)$: Spin excitation spectrum and $\omega_c$ along $q_x=q_y$.}
\label{width}
\end{figure}

We consider next the resonance peak in the underdoped YBa$_2$Cu$_3$O$_{6+x}$ compounds; our results are summarized in
 inset $(a)$ of Fig.~\ref{width}. Fong {\it et al.}~find
 that $\omega_{res}$ decreases to $\omega_{res}=25$ meV in YBa$_2$Cu$_3$O$_{6.5}$ while  the integrated intensity of the peak increases from  I$_{int} \approx 1.1 \, \mu_B^2$ in  YBa$_2$Cu$_3$O$_7$ to I$_{int} \approx  2.6 \, \mu_B^2$ in  YBa$_2$Cu$_3$O$_{6.5}$. In our model, the doping dependence of 
$\omega_{res}=\Delta_{sw}=c_{sw}/\xi$ is determined by changes in $c_{sw}$ and $\xi$. From an analysis of NMR \cite{Zha96} and INS experiments \cite{Morr98}, we know that  the correlation length at $T_c$ increases from  $\xi \approx 2.2$ in YBa$_2$Cu$_3$O$_7$ to $\xi \approx 6$ in  YBa$_2$Cu$_3$O$_{6.5}$. This change in $\xi$ more than compensates any increase of $c_{sw}$ as the doping is reduced and so brings about a decrease in $\Delta_{sw}$.
Furthermore, Bourges {\it et al.} \cite{Bou97} find a normal state spin-gap in YBa$_2$Cu$_3$O$_{6.5}$, $\Delta_{sw} \approx 23$ meV \cite{Bou97}, in agreement with the position of the resonance peak in the superconducting state reported by Fong {\it et al.} \cite{Fong97}. 
The integrated intensity I$_{int}$ of the resonance peak in the limit 
Im$\, \Pi_{odd} \ll 1/\alpha\xi^2$, i.e., for small spin damping, is given by 
$ {\rm I}_{int} = \alpha \xi^2 \Delta_{sw} \pi /4 $.
For YBa$_2$Cu$_3$O$_7$, $\alpha=15 \mu_B^2/eV$ which yields I$_{int}= (2.3\pm 0.4)\mu_B^2$. For YBa$_2$Cu$_3$O$_{6.5}$, we find from our analysis of INS data in the normal state $\alpha \approx 9 \mu_B^2/eV$, and thus I$_{int}= (6.2\pm 1.0) \mu_B^2$.  
Both values for I$_{int}$ are a factor of  2 larger than the values measured experimentally by 
Keimer {\it et al.} \cite{Kei97}; however, the relative increase of  $I_{int}$ on going from  YBa$_2$Cu$_3$O$_7$ to YBa$_2$Cu$_3$O$_{6.5}$ is in  quantitative agreement with their results. 

As $2\Delta_{SC}$ and $\Delta_{sw}$ become well separated in the underdoped compounds \cite{arpes,Ren98}, we predict a resonance peak for all momenta in a region around ${\bf Q}$ which is determined by $\omega_q< \omega_c$ (see inset $(b)$ of Fig.~\ref{width}).
In Fig.~\ref{width} we show how the resonance peak shifts to higher frequencies when one moves in momentum space from   
${\bf Q}=(\pi,\pi)$ to ${\bf q}=0.97(\pi,\pi)$.
Due to the limited momentum resolution of INS experiments we do not expect that these peaks can be resolved; this implies that one should only observe one broad peak which extends from $\Delta_{sw}$ up to $\omega_c$. This broadening of the peak width is just what is observed by Fong {\it et al.} \cite{Fong97} in YBa$_2$Cu$_3$O$_{6.5}$ and YBa$_2$Cu$_3$O$_{6.7}$.

For the underdoped compounds, two groups find a precursor of the resonance peak in the normal state below $T_* \approx 200$ K, in the so-called 
strong pseudo-gap region \cite{Fong97,Bou97}. At the same time, ARPES experiments find that below $T_*$, the low-frequency spectral weight of the "hot" fermionic quasi-particles around $(0,\pi)$ is suppressed 
(the "leading-edge gap") \cite{arpes}, which immediately  leads to a decrease of the spin-damping at ${\bf Q}=(\pi,\pi)$ and an enhancement of $\chi''_{odd}$. 
Our model thus predicts an enhancement of $\chi''_{odd}$ around $\omega_{res}$ in the normal state of those superconductors for which a leading edge gap occurs below $T_*$.

It follows from the above analysis, that a resonance peak should occur in all cuprate superconductors for which the condition $\Delta_{sw} < 2 \Delta_{SC}$ is satisfied. 
This implies that the resonance peak is {\it not} directly related to the bilayer structure of YBa$_2$Cu$_3$O$_{6+x}$.
In La$_{2-x}$Sr$_x$CuO$_4$, where $\Delta_{SC} \approx 9$ meV and $\Delta_{sw}> 25$ meV \cite{Mas98}, the above condition is not satisfied and consequently, no resonance peak is to be expected. On the other hand, for the single-layer material HgBa$_2$CuO$_{4+\delta}$ with an optimum $T_c=95K$, we expect $\Delta_{SC}$ to be comparable to that found in 
YBa$_2$Cu$_3$O$_7$, so that it is a good candidate for the observation of a resonance peak.  

Our model also provides a natural explanation for the results of  recent INS experiments on Zn-doped YBa$_2$Cu$_3$O$_{6.97}$ \cite{Sid96}. 
A Zn concentration of $2\%$ which suppresses $T_c$ to $T_c=69K$ 
completely destroys the resonance peak in the superconducting state, and generates a significant amount of spectral weight in $\chi''$ at low frequencies. Since $\Delta_{sw} \approx 2 \Delta_{SC}$ in this material,  to the extent that $\Delta_{sw}$ is not markedly influenced by Zn, a quite modest Zn-induced decrease in $\Delta_{SC}$ will render the resonance unobservable in the superconducting state. For the underdoped compounds, on the other hand, since $\Delta_{sw} $ and $2 \Delta_{SC}$ are more clearly separated, a larger amount of Zn would be required to suppress the resonance.
The appearance of low-frequency spectral weight in $\chi''$ shows that a considerable spin-damping persists in the superconducting state, as might be expected from Zn-induced changes in the low-frequency spectral weight in the single-particle spectrum.  

In summary we have shown that a spin-wave model provides a natural explanation for the  doping and temperature dependence of the existing INS experiments on the appearance of a resonance peak in the superconducting state  and of a resonance feature in the normal state of underdoped systems. We proposed necessary conditions for the observation of the peaks and resonance features, and so explain the failure to observe either of them in   La$_{2-x}$Sr$_x$CuO$_4$.
We argue that the occurrence of the resonance peak is not  directly related to the bilayer structure of  YBa$_2$Cu$_3$O$_{6+x}$, and predict that it should also be observable in single-layer cuprates with a sufficiently large superconducting gap.

We would like to thank P. Bourges, A.V. Chubukov, P. Dai, T. Fong, B. Keimer, B. Lake, T. Mason, A. Millis, H. Mook,  and J. Schmalian for valuable discussions. This work has been supported in part by the Science and Technology Center for Superconductivity through NSF-grant DMR91-20000.


\begin{thebibliography}{99}
\bibitem{Exp1} J. Rossat-Mignod {\it et al.} Physica C {\bf 185-189}, 86 (1991); H. A. Mook {\it et al.} Phys. Rev. Lett. {\bf 70}, 3490 (1993); H.F. Fong {\it et al.}, Phys. Rev. Lett. {\bf 75}, 316 (1995); P. Bourges {\it et al.} Phys. Rev. 
B {\bf 53}, 876 (1996). 
\bibitem{Dai96} P. Dai {\it et al.}, Phys. Rev. Lett. {\bf 77}, 5425 (1996).
\bibitem{Dai98} P. Dai {\it et al.},Phys. Rev. Lett. {\bf 80}, 1738 (1998).
\bibitem{Fong97} H.F. Fong {\it et al.}, Phys. Rev. Lett. {\bf 78}, 713 (1997).
\bibitem{Kei97} B. Keimer {\it et al.}, preprint, cond-mat 9705103. 
\bibitem{Hay97} S. Hayden {\it et al.}, preprint, cond-mat 9710181.
\bibitem{Bou97} P. Bourges {\it et al.}, Phys. Rev. B {\bf 56}, R11439 (1997); P. Bourges, preprint.
\bibitem{Fong96} H.F. Fong {\it et al.} Phys. Rev. B {\bf 54}, 6708 (1996).
\bibitem{final} I.I. Mazin and V.M. Yakovenko, Phys. Rev. Lett. {\bf 75}, 4134 (1995); D.Z. Liu, Y. Zha, and K. Levin, Phys. Rev. Lett. {\bf 75}, 4130 (1995); A. Millis and H. Monien, Phys. Rev. B {\bf 54}, 16172 (1996).
\bibitem{anom} N. Bulut and D. Scalapino, Phys. Rev. B {\bf 53}, 5149 (1996); G. Blumberg, B.P. Stojkovic, and M.V. Klein, Phys. Rev. B {\bf 52}, R15741 (1996).
\bibitem{Yin97} L. Yin, S. Chakravarty, P.W. Anderson, Phys. Rev. Lett. {\bf 78}, 3559 (1997).
\bibitem{SO5} E. Demler and S. C. Zhang, Phys. Rev. Lett. {\bf 75}, 4126 (1995); J. Brinckmann and P. A. Lee, preprint, cond-mat 9710065.
\bibitem{Bar95} V. Barzykin and D. Pines, Phys. Rev. B {\bf 52}, 13585 (1995).
\bibitem{arpes} A.G. Loeser {\it et al.}, Science {\bf 273}, 325 (1996); H. Ding 
 {\it et al.}, Nature {\bf 382}, 51 (1996).
\bibitem{Ren98} Ch. Renner {\it et al.}, Phys. Rev. Lett. {\bf 80}, 149 (1998).
\bibitem{com1} The exact value of $\omega_c \approx 2\Delta_{SC} $ depends on bandstructure parameters; since it is not important for our 
scenario,  for simplicity we take $\omega_c=2\Delta_{SC}$. 
\bibitem{sfmodel}  P. Monthoux and D. Pines, Phys. Rev B {\bf 47}, 6069 (1993).
\bibitem{Chu94} A.V. Chubukov, S. Sachdev, and J. Ye, Phys. Rev. B {\bf 49}, 11919 (1994). 
\bibitem{Morr98} D.K. Morr and D. Pines, in preparation.
\bibitem{Zha96} Y. Zha, V. Barzykin and D. Pines, Phys. Rev. B {\bf 54}, 7561
(1996).
\bibitem{Mag96} I. Maggio-Aprile {\it et al}, J. Low Temp. Phys. {\bf 105}, 1129 (1996).
\bibitem{Chu97} A.V. Chubukov and D.K. Morr, Phys. Rep. {\bf 288}, 355 (1997).
\bibitem{Mil98} C. Milling, private communication.
\bibitem{Mas98} T. Mason, private communication.
\bibitem{Sid96} Y. Sidis {\it et al.}, Phys. Rev B {\bf 53}, 6811 (1996).
\end{thebibliography}
 \end{document}